\documentclass[12pt]{article}
\usepackage{amsmath,amssymb,amsfonts, epsfig}
\usepackage{color}

\makeatletter

\@addtoreset{equation}{section}
\makeatother

\begin{document}
\title{
\begin{flushright}
\ \\*[-80pt] 
\begin{minipage}{0.2\linewidth}
\normalsize
%arXiv:YYMM.NNNN \\
KUNS-2462  \\
TU-946 \\
MISC-2013-08  \\*[50pt]
\end{minipage}
\end{flushright}
{\Large \bf 
CKM matrix and flavor symmetries
\\*[20pt]}}

\author{
Takeshi~Araki,$^{1}$\ 
Hiroyuki~Ishida,$^{2}$\
Hajime~Ishimori, $^{3}$ \\
Tatsuo~Kobayashi,$^{3}$ \
Atsushi~Ogasahara $^{3}$  \\
\\*[20pt]
\centerline{
\begin{minipage}{\linewidth}
\begin{center}
$^1${\it \normalsize
Maskawa Institute, Kyoto Sangyo University, Kyoto 603-8555, Japan} \\
$^2${\it \normalsize
Department of Physics, Tohoku University, Sendai 980-8578, Japan}\\
$^3${\it \normalsize 
Department of Physics, Kyoto University, 
Kyoto 606-8502, Japan} \\
\end{center}
\end{minipage}}
\\*[50pt]}
\vskip 2 cm
\date{\small
\centerline{ \bf Abstract}
\begin{minipage}{0.9\linewidth}
\medskip
Following the way proposed recently by Hernandez and Smirnov, 
we seek possible residual symmetries in the quark sector with a focus on the von Dyck groups.
We begin with two extreme cases in which both $\theta_{13}$ and $\theta_{23}$ or only $\theta_{13}$ are set to zero.
Then, cases where all the Cabibbo-Kobayashi-Maskawa parameters are allowed to take nonzero
values are explored.
The $Z_7$ symmetry is favorable to realize only the Cabibbo angle. 
On the other hand, larger groups are necessary in order to be consistent with all the mixing parameters.
Possibilities of embedding the obtained residual symmetries into the $\Delta(6N^2)$ series are also briefly discussed.
\end{minipage}
}

\begin{titlepage}
\maketitle
\thispagestyle{empty}
\end{titlepage}

\section{INTRODUCTION}

Theoretical understanding of the observed quark and lepton mixings 
is one of the longstanding issues in the standard model (SM).
Indeed, various attempts to understand them have been done 
by imposing symmetries, using some dynamics or assuming 
extra dimensions.

Neutrino oscillation experiments show 
large mixing angles in the lepton sector except $\theta^\ell_{13}$~\cite{pdg}.
In the approximation with $\theta^\ell_{13} \approx 0$, 
the tribimaximal mixing matrix is a quite interesting 
ansatz for the lepton sector~\cite{Harrison:2002er}.
The tribimaximal mixing matrix has certain symmetries.
Then, a number of studies have been carried out to 
derive it by using non-Abelian discrete flavor symmetries 
(see for review Refs.~\cite{Altarelli:2010gt,Ishimori:2010au,King:2013eh}.)
In those studies, first a non-Abelian flavor symmetry $G^{(\ell)}_f$ 
for the lepton sector is assumed.
Then, such a symmetry is broken to $G_\ell$ ($G_\nu$) in the mass terms 
of the charged lepton (neutrino) sector.
It was also found that inherent symmetries
 $G_\nu = Z_2 \times Z_2$ and  $G_\ell=Z_3$  
in a certain basis are important to derive the tribimaximal mixing
matrix~\cite{resdS4}.

Recent neutrino experiments show that $\theta^\ell_{13} \neq 0$~\cite{rct,acl}.
However, the above approach to use flavor symmetries  
is still interesting to derive experimental values of lepton mixing angles (see e.g. Ref.~\cite{Ge:2011qn}), 
although we need some modifications.
For example, $G_\ell$ was often extended from $Z_3$ to $Z_m$.

Hernandez and Smirnov developed a model-independent way of 
screening out possible flavor symmetries in the lepton sector to 
derive experimental values~\cite{Hernandez:2012ra}.(See also Ref.~\cite{Hu:2012ei}.)
They figured out necessary conditions for ensuring 
that the inherent symmetries $G_\nu = Z_2( \times Z_2)$ and  $G_\ell=Z_m$
can be embedded into a single discrete group.
In particular, the lepton mixing angles are written interestingly in terms of 
a small number of integers by requiring that 
the product $g$ between the $Z_m$ element in $G_\ell$ and the $Z_2$ in $G_\nu$
should satisfy $g^p=1$, that is, $g$ is a $Z_p$ element.

The mixing angles in the quark sector, the Cabibbo-Kobayashi-Maskawa 
(CKM) matrix, 
are another issue to study.
These patterns are quite different from those in the lepton sector.
These mixing angles except the Cabibbo angle are rather small.
There is no ansatz to figure out underlying symmetries 
behind the CKM matrix like the tribimaximal mixing ansatz.
Here we study systematically the symmetries of the CKM matrix following 
Hernandez-Smirnov's analysis on the lepton sector.
We assume that the flavor symmetry $G_f^{(q)}$ in the quark sector 
is broken  to $G_u$ ($G_d$) in the mass terms 
of the up-quark (down-quark) sector, where 
$G_u= Z_n$ and $G_d=Z_m$.
Then, following Hernandez and Smirnov, 
we assume that the product $g$ between the $Z_m$ element in $G_d$ and
the $Z_n$ in  $G_u$
should satisfy $g^p=1$, that is, $g$ is a $Z_p$ element. 
That leads to constraints on three angles and one phase in the CKM 
matrix depending on $m,~n$ and $p$.
Then, we would figure out what Abelian symmetries are 
important to realize the CKM matrix as 
$Z_2 \times Z_2$ and $Z_3$ symmetries are important 
to realize the tribimaximal mixing matrix.
Such an analysis is useful to investigate candidates 
for the quark flavor symmetry $G_f^{(q)}$, which includes 
$Z_n$, $Z_m$ and $Z_p$.

This paper is organized as follows. 
In Sec. \ref{sec:application}, we apply Hernandez-Smirnov's analysis to the quark sector and derive conditions on the CKM parameters.
For some special cases, we consider the conditions and 
derive possible residual symmetries in Sec. \ref{sec:quark_mixing}.
Possibilities of embedding the residual symmetries into the $\Delta(6N^2)$ groups are discussed in Sec. \ref{sec:delta}.
We summarize our discussions in Sec. \ref{sec:summary}.

\medskip

\section{APPLICATION TO THE QUARK SECTOR} \label{sec:application}

In this section, following the way proposed by Hernandez and Smirnov, we apply their method to the quark sector.
Of the SM Lagrangian, only the quark mass terms are relevant to our discussion, which are written by
\begin{align}
-{\cal L}\,=\,\bar{U}_R {\hat M}_U U_L + \bar{D}_R {M}_D D_L+H.c. \,,
\label{eq:L}
\end{align}
where $U_{L,\,R}\equiv(u,\, c,\, t)_{L,\,R}^T$ and 
$D_{L,\, R}\equiv(d,\,s,\,b)_{L,\, R}^T$ are the up- and down-type quarks, respectively, and 
${\hat M}_U \equiv \text{diag}\{m_u,\,m_c,\,m_t\}$ is the diagonalized up quark mass matrix. 
The down quark mass matrix, in this basis, can be taken to be
\begin{align}
M_{D}\,=\, V_{\rm CKM}^{} {\hat M}_{D}V_\text{CKM}^\dag \,,
\end{align}
where $V_\text{CKM}$ is the CKM matrix, 
and ${\hat M}_{D} \equiv \text{diag}\{m_d,\,m_s,\,m_b\}$. 
The CKM matrix is determined by the four parameters, $(\theta_{12},\, \theta_{13}, \,\theta_{23}, \,\delta)$, and we obey the standard parametrization:
\begin{equation}
V_{\text{CKM}}\,=\,\begin{pmatrix}
c_{12}c_{13} \,&&\, s_{12}c_{13} \,&&\, s_{13}e^{-i\delta} \\
-s_{12}c_{23}-c_{12}s_{23}s_{13}e^{i\delta} \,&&\, c_{12}c_{23}-s_{12}s_{23}s_{13}e^{i\delta} \,&&\, s_{23}c_{13} \\
s_{12}s_{23}-c_{12}c_{23}s_{13}e^{i\delta} \,&&\, -c_{12}s_{23}-s_{12}c_{23}s_{13}e^{i\delta} \,&&\, c_{23}c_{13} 
\end{pmatrix}\,,
\end{equation}
where $s_{ij}$ and $c_{ij}$ represent $\sin\theta_{ij}$ and $\cos\theta_{ij}$, respectively. 
In what follows, we will use the following central values given in Ref.~\cite{pdg}:
\begin{eqnarray}
(\sin\theta_{12},\,\sin\theta_{23},\,\sin\theta_{13},\, \cos\delta )=(0.225,\, 0.0412,\, 0.00341,\, 0.355)\,.
\end{eqnarray}

In general, each quark mass term inherently and maximally respects three $U(1)$ symmetries corresponding to three generations.
For instance, in the basis of Eq.~\eqref{eq:L}, the down quark mass term is invariant under transformations
\begin{equation}
D_R \to\, S_D D_R \,,\qquad
D_L \to\, S_D D_L \,,
\end{equation}
where 
\begin{equation}
S_0\,\equiv\, \text{diag} \{e^{i\phi_d},\,e^{i\phi_s},\,e^{i\phi_b}\}\,,\qquad
S_D \,=\, V_\text{CKM}S_0 V^\dagger_\text{CKM}\,. \label{eq:s_D}
\end{equation}
We regard $S_D$ as an element of a residual symmetry $G_d$ arising from spontaneous breaking of a full flavor symmetry $G_f^{(q)}$ at a high energy scale.
In this work, for simplicity, we assume that $G_f^{(q)}$ and $G_d$ are discrete subgroups 
of $SU(3)$ and that $G_d$ is $Z_m$. Then $S_D$ satisfies 
\begin{equation}
S_D^m\,=\,{\mbox{1}\hspace{-0.25em}\mbox{l}}\,.
\end{equation}
The phases in $S_D$ are described as
\begin{gather}
\phi_d\,\equiv\,2\pi \frac{k_d}{m}\,,\qquad 
\phi_s\,\equiv\,2\pi \frac{k_s}{m}\,,\qquad 
\phi_b\,\equiv\,2\pi \frac{k_b}{m}\,,
\label{eq:phi_dsb=0}
\end{gather}
where $(k_d,\,k_s,\,k_b)$ represent $Z_m$ charges and are restricted to be
\begin{align}
k_d+k_s+k_b\,\equiv\, 0\qquad \text{mod}~m \,, \label{eq:k-down}
\end{align}
because of  the assumption $G_d \subset SU(3)$.

Similarly, in the up quark mass term, we postulate $G_u = {Z}_n$ and define $Z_n$ transformations as 
\begin{align}
U_L \to\, T U_L \,,\qquad 
U_R \to\, T U_R\,, 
\end{align}
where
\begin{gather}
T \,\equiv\, \text{diag}\{ e^{i\phi_u},\, e^{i\phi_c},\, e^{i\phi_t}\}\,,\\
T^n \,=\, {\mbox{1}\hspace{-0.25em}\mbox{l}}\,, \\
\phi_u \,\equiv\, 2\pi \frac{k_u}{n} \,,\qquad
\phi_c \,\equiv\, 2\pi \frac{k_c}{n} \,,\qquad 
\phi_t \,\equiv\, 2\pi \frac{k_t}{n} \,, 
\end{gather}
and, from the assumption $G_u \subset SU(3)$,\footnote{
Note that since Eqs.~\eqref{eq:k-down} and \eqref{eq:k-up} lead to 
$\det S_0= \det S_D = \det T= 1$, 
these $Z_m$ and $Z_n$ symmetries are anomaly free (see
Ref.~\cite{Araki:2008ek} and references therein).}
\begin{equation}
k_u + k_c + k_t \,\equiv\, 0\qquad\text{mod}~n\,. \label{eq:k-up}
\end{equation}

In order for both $G_u$ and $G_d$ to be residual symmetries of $G_f^{(q)}$, products of $S_D$ and $T$ must also belong to $G_f^{(q)}$ and have a finite order.
Hence, we here introduce a new element
\begin{align}
W \,\equiv\, S_D T \,,\label{eq:W}
\end{align}
and require that $W$ is an element of $Z_p$, leading to
\begin{align}
W^p \,=\, (S_D T)^p \,=\, {\mbox{1}\hspace{-0.25em}\mbox{l}}\,.
\end{align}
Thus, in total, $G_f^{(q)}$ should contain three elements, 
$S_D$,  $T$ and $W$, satisfying
\begin{align}
S_D^m\,=\, T^n\,=\,W^p\,=\,{\mbox{1}\hspace{-0.25em}\mbox{l}}\,,
\end{align}
and groups composed of such elements are called von Dyck group.
The groups have the finite number of elements if $1/n+1/m+1/p > 1$, whereas it is infinite in the case of $1/n+1/m+1/p \le 1$.

The requirement of $W^p={\mbox{1}\hspace{-0.25em}\mbox{l}}$ gives two constraints on the CKM matrix elements.
In order to demonstrate that, let us consider the trace of $W$. 
On one hand, $\text{tr}[W]$ can directly be written down with $W=V_{\rm CKM}S_0 V_{\rm CKM}^\dag T$.
On the other hand, $\text{tr}[W]$ is equal to a sum of the eigenvalues of $W$. 
{}From $W^p={\mbox{1}\hspace{-0.25em}\mbox{l}}$, the sum, $a$, is given by
\begin{gather}
a\,=\, 
e^{2\pi i \frac{q_1}{p}}+e^{2\pi i \frac{q_2}{p}}+e^{2\pi i \frac{q_3}{p}}\,,\qquad
q_1+q_2+q_3\,\equiv\, 0\qquad\text{mod}~p\,,
\end{gather} 
where $(q_1,\,q_2,\,q_3)$ represent $Z_p$ charges. 
As a result, one obtains
\begin{align}
tr[W]\,=\,&e^{i\phi_u}\left[ |V_{ud}|^2 e^{i \phi_d}+ |V_{us}|^2 e^{i \phi_s}+ |V_{ub}|^2 e^{i \phi_b}\right] \nonumber \\
&+e^{i\phi_c}\left[ |V_{cd}|^2 e^{i \phi_d}+ |V_{cs}|^2 e^{i \phi_s}+ |V_{cb}|^2 e^{i \phi_b}\right] \nonumber \\
&+e^{i\phi_t}\left[ |V_{td}|^2 e^{i \phi_d}+ |V_{ts}|^2 e^{i \phi_s}+ |V_{tb}|^2 e^{i \phi_b}\right] \,=\, a \,,
\label{eq:trW=a}
\end{align}
where $V_{ij}$ stands for the $ij$ element of the CKM matrix.
Since $a$ is a complex parameter, Eq.~\eqref{eq:trW=a} generally yields two constraints on 
the CKM matrix elements with a fixed integer set 
$(n,\,k_u,\,k_c,\,m,\,k_d,\,k_s,\,p,\,q_1,\,q_2)$.
Since the CKM matrix is characterized by the 
four parameters $(\theta_{12},\, \theta_{13}, \,\theta_{23}, \,\delta)$, 
the above two constraints cannot determine the whole CKM matrix. 
However, by setting some of the CKM parameters by hand, 
one can predict the remaining CKM parameters. 
In the following section, under several cases, we systematically 
search the integer sets $(n,\,k_u,\,k_c,\,m,\,k_d,\,k_s,\,p,\,q_1,\,q_2)$,
which predict the experimental data. 

\medskip

\section{QUARK MIXING}\label{sec:quark_mixing}

\subsection{$\theta_{13}=\theta_{23}=0$ case}\label{sec:cabibbo}

By fixing at least two of the four CKM parameters by hand, one can predict the remaining CKM parameters with Eq.~\eqref{eq:trW=a} for given $(n,\,k_u,\,k_c,\,m,\,k_d,\,k_s,\,p,\,q_1,\,q_2)$.
Since $\theta_{13}$ and $\theta_{23}$ are much smaller than $\theta_{12}$, they are often taken to be vanishing at zeroth order in some flavor models.
We here adopt the same stance and concentrate on predicting the Cabibbo angle $\theta_{12}$.
Note that we do not aim at precisely reproducing the Cabibbo angle and others.
Small perturbations are expected to occur, because a breaking scale of the full flavor symmetry is supposed to be very high.

In the case of $\theta_{13}=\theta_{23}=0$, the Dirac phase $\delta$ also disappears from the CKM matrix, then Eq.~\eqref{eq:trW=a} is reduced to be
\begin{align}
e^{i(\phi_u+\phi_d)}+e^{i(\phi_c+\phi_s)}+e^{i(\phi_t+\phi_b)}-s_{12}^2(e^{i\phi_u}-e^{i\phi_c})(e^{i\phi_d}-e^{i\phi_s})\,=\,a\,.
\label{eq:trW_cab}
\end{align}
This yields two conditions on $\theta_{12}$:
\begin{gather}
\sin^2\theta_{12}\,=\,\frac{-\text{Re}[a]+\cos(\phi_b+\phi_t)+\cos(\phi_d+\phi_u)+\cos(\phi_s+\phi_c)}{\cos(\phi_d+\phi_u)-\cos(\phi_d+\phi_c)
-\cos(\phi_s+\phi_u)+\cos(\phi_s+\phi_c)}\,,
\label{eq:sin12_re}
\end{gather}
 from the real part while
\begin{gather}
\sin^2\theta_{12}\,=\,\frac{-\text{Im}[a]+\sin(\phi_b+\phi_t)+\sin(\phi_d+\phi_u)+\sin(\phi_s+\phi_c)}{\sin(\phi_d+\phi_u)-\sin(\phi_d+\phi_c)
-\sin(\phi_s+\phi_u)+\sin(\phi_s+\phi_c)}\,,
\label{eq:sin12_im}
\end{gather}
from the imaginary part.
In general, these conditions are independent 
of each other and must simultaneously be satisfied.
However, in some cases, restrictions on $\theta_{12}$ become reduced. 
For example, when the denominator of Eq.~\eqref{eq:sin12_re} [or Eq.~\eqref{eq:sin12_im}] is zero, 
the real (or the imaginary) part of Eq.~\eqref{eq:trW_cab} becomes free from $\theta_{12}$, which 
implies that Eq.~\eqref{eq:sin12_re} [or Eq.~\eqref{eq:sin12_im}] gives no constraint on $\theta_{12}$. 
Furthermore, in certain cases, both conditions, Eqs.~\eqref{eq:sin12_re} and~\eqref{eq:sin12_im}, provide no constraints on $\theta_{12}$, e.g. the denominators are vanishing in both conditions; we will omit such cases in what follows.

\begin{figure}[htbp]
\begin{center}
\includegraphics[width=10cm]{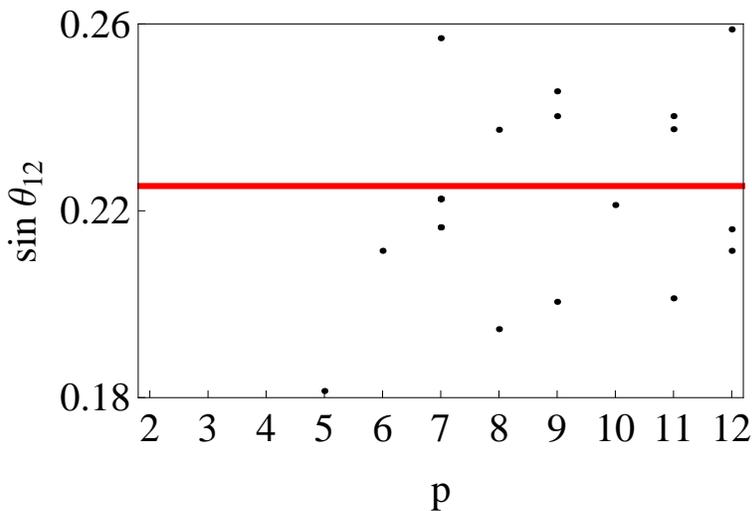}\medskip
\caption{Distribution of $\sin\theta_{12}$ as a function of $p$ up to $(n,\, m,\,p)=(6,\,6,\,12)$ while setting $\theta_{23}=\theta_{13}=0$. The bold line is the experimental value, $\sin\theta_{12}=0.225$.}
\label{fig:cab}
\end{center}
\end{figure}

We perform a numerical search for a possible integer set 
$(n,\,k_u,\,k_c,\,m,\,k_d,\,k_s,\,p,\,q_1,\,q_2)$ up to $(n,\, m,\,p)=(6,\,6,\,12)$.\footnote{When $n=2$ or $m=2$, 
one needs to introduce an additional $Z_2$ symmetry 
to keep the mass matrix diagonal in their diagonal basis. 
The additional $Z_2$ can be chosen not to affect $\theta_{12}$.}
Figure \ref{fig:cab} shows $\sin\theta_{12}$ near the experimental value as a function of $p$.
We also list the seven nearest solutions in Table \ref{table:cab} and find that all the solutions are within the region of $0.21 < \sin\theta_{12} < 0.24$.
In the table, only the case of $(n,\,m,\,p)=(2,\,2,\,7)$ can be a finite von Dyck group, 
which is $D_7$ and discussed in Ref.~\cite{D7}, while all the other cases compose infinite von Dyck groups.
Some of the latter cases may be embedded into finite subgroups of infinite von Dyck 
groups \cite{deAdelhartToorop:2011re, Hernandez:2012ra}.

Let us study more on the case with $n=m=2$.
We consider  a special case with $n=m=2$, $k_u=k_s=q_1=0$, and $k_c=k_d=1$.
Then, the mixing angle in Eq.~\eqref{eq:sin12_re} 
can be expressed by a simple form of 
$\sin^2\theta_{12}=\cos^2(\pi q_2/p)$. 
The mixing angle in this form can be derived by the 
finite group $\Delta (6N^2)$ as discussed in Sec.~\ref{sec:delta}. 
Experimentally, the center value $\sin \theta_{12}$ is $0.225$, 
so we need $q_2/p= 0.428$. 
Also, the range $\sin \theta_{12} = [0.224,0.226] $ 
corresponds to  $q_2/p= [0.427,0.428]$.
To express this value approximately by integer sets, 
the combination of $p=7$ and $q_2=3$ is the 
simplest, i.e. $3/7=0.4286$, 
 as it appeared in our numerical analysis. 
For $p < 89$, there is no interger to fit better than $p=7$ 
except $(p,q_2)=(7r,3r)$.
Taking greater number for $p$, we can better fit 
to the experimental value, 
for instance $(p,q_2)=(89,38),~(96,41),~(103,44)$, i.e. 
$38/89=0.4270$, $41/96=0.4271$, $44/103=0.4272$.
Thus, $p=7$ is much simpler to fit the experimental value in this
scheme and it seems that the $Z_{7r}$ symmetry is favorable.

\medskip

\begin{table}[htbp]
\begin{center}
\begin{align}
\begin{array}{c|ccc|cccc|ccc}\hline\hline
\sin\theta_{12}&n&m & p & k_u&k_c&k_d&k_s& q_1 & q_2 & (\text{Re}[a],\, \text{Im}[a])\\ \hline
 0.223 &  2&2&7&0&1&1&0&0&3& ( -0.802,\,  0)\\
 0.211 & 2&5&6&0&1&1&4&1&2& (-1.00,\, 1.73)\\
0.216&3&5&7&1&2&1&4&0&3& (-0.802,\, 0)\\
0.216&3&5&12&0&1&3&2&7&8&(-1.37,\, -2.37) \\
0.237&3&6&11&0&1&1&3&0&2&(1.83,\, 0) \\
0.221&4&4&10&0&1&1&0&2&3&(-1.00,\, 1.90)\\
0.237&5&5&8&0&2&3&0&0&3&(-0.414,\,0)\\ \hline
\end{array}\nonumber
\end{align}\medskip
\caption{While setting $\theta_{23}=\theta_{13}=0$ and up to $(n,\, m,\,p)=(6,\,6,\,12)$, we pick out seven solutions 
close to the experimental value; all of them are within the range of $0.21 < \sin\theta_{12} < 0.24$.
Integer sets which give the same $\sin\theta_{12}$ displayed here are omitted.
}
\label{table:cab}
\end{center}
\end{table}

\medskip

\subsection{$\theta_{13}=0$ case}\label{sec:13=0}
Next, we lift the restriction of $\theta_{23}=0$.
Then, the expression of $\text{tr}[W]$ is given as follows:
\begin{align}
&
e^{i(\phi_u+\phi_d)}+e^{i(\phi_c+\phi_s)}+e^{i(\phi_t+\phi_b)}-s_{12}^2(e^{i\phi_u}-e^{i\phi_c})(e^{i\phi_d}-e^{i\phi_s})
\nonumber \\
&
-s_{23}^2(e^{i\phi_c}-e^{i\phi_t})(e^{i\phi_s}-e^{i\phi_b}) - s_{12}^2s_{23}^2(e^{i\phi_c}-e^{i\phi_t})(e^{i\phi_d}-e^{i\phi_s})\,=\,a \,.
\label{eq:trW_(13=0)}
\end{align}
As in Sec. \ref{sec:cabibbo}, we focus only on solutions in 
which $\theta_{12}$ and $\theta_{23}$ are uniquely determined. 
In addition, we here omit solutions predicting $\theta_{23}=0$ since they are included in the results of Sec. \ref{sec:cabibbo}.\footnote{
However, the opposite is not always true.
For example, the solution of $(n,\,m,\,p)=(2,\,2,\,7)$ in Table \ref{table:cab} 
does not derive $\theta_{23}=0$ in Eq.~\eqref{eq:trW_(13=0)}.
This is because the terms associated with $\theta_{23}$ are dropped from Eq.~\eqref{eq:trW_(13=0)}, and thus $\theta_{23}$ is not determined.
}

\begin{figure}[htbp]
\begin{center}
\includegraphics[width=10cm]{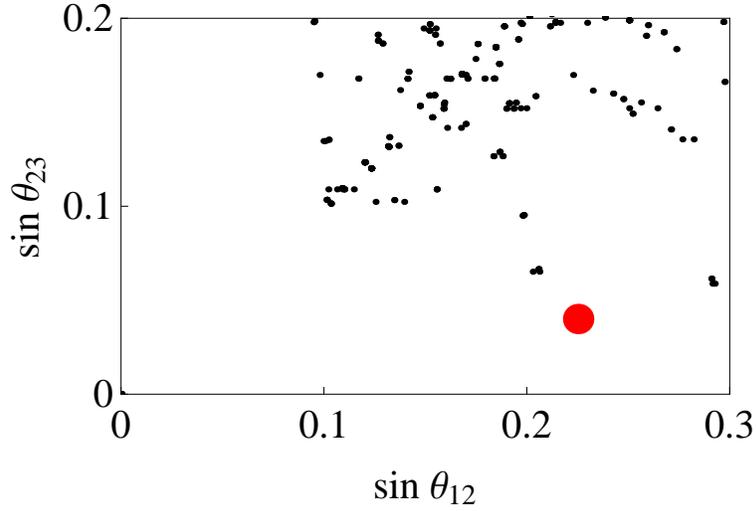}\medskip
\caption{Distribution of $\sin\theta_{23}$ as a function of  $\sin\theta_{12}$ up to $(n,\,m,\,p)=(6,\,6,\,12)$ while setting $\theta_{13}=0$. 
Their experimental values are depicted as the large red point at $(\sin\theta_{12},\sin\theta_{23})=(0.225,\, 0.0412)$.}
\label{fig:s13is0}
\end{center}
\end{figure}

Similar to Sec. \ref{sec:cabibbo}, we search for possible integer sets up to $(n,\,m,\,p)=(6,\,6,\,12)$. 
We plot $\sin\theta_{23}$ as a function of $\sin\theta_{12}$ in Fig. \ref{fig:s13is0} 
and pick out seven solutions close to the experimental values in Table \ref{table:s13is0}. 
All the solutions in Table \ref{table:s13is0} satisfy $0.15<\sin\theta_{12}<0.30$ 
and $0.05<\sin\theta_{23}<0.10$. 
{}From Fig. \ref{fig:s13is0}, it can be seen that $\sin\theta_{12}$ can be near the experimental value, 
whereas most of the obtained $\sin\theta_{23}$ are much larger than its experimental value.
Moreover, solutions which are marginally consistent with the experimental 
value of $\sin\theta_{23}$ are obtained for relatively large $(n,\,m,\,p)$.
These tendencies may suggest that larger $(n,\,m,\,p)$ are necessary 
for reproducing a realistic $\sin\theta_{23}$ as well as $\sin\theta_{12}$.

\medskip

\begin{table}[htbp]
\begin{center}
\begin{align}
\begin{array}{c|ccc|cccc|ccc}\hline\hline
(\sin\theta_{12},\,\sin\theta_{23})&n&m & p & k_u&k_c&k_d&k_s& q_1&q_2&(\text{Re}[a],\, \text{Im}[a])\\ \hline
(0.206,\, 0.0666) &4&5& 11&0& 1& 4& 2&6&7& (-1.20,\, -1.95)\\
(0.206,\, 0.0652)&5& 4&11& 1& 3& 0& 3&2&4& (-1.20,\, 1.95) \\
(0.203,\,0.0651)&5&4&11&4&2&0&1&6&7&(-1.20,\, -1.95)\\
(0.291,\, 0.0588)&5& 6&11&  1& 3& 0& 5&2&4& (-1.20,\, 1.95)\\
(0.198,\, 0.0950)& 5& 6& 11& 2& 0& 2& 1&1&2& (1.11,\, 0.460)\\ 
(0.291,\, 0.0615)&6&5&11&0&1&4&2&6&7&(-1.20,\, -1.95)\\
(0.198,\, 0.0953) &6 &5 &11&2&1 &2 &0 &1&2& (1.11,\, 0.460) \\\hline
\end{array}\nonumber
\end{align}\medskip
\caption{While setting $\theta_{13}=0$ and up to $(n,\,m,\,p)=(6,\,6,\,12)$, we pick out seven solutions close to the experimental values. 
The solutions are all within $0.15<\sin\theta_{12}<0.30$ and $0.05<\sin\theta_{23}<0.10$.
Solutions predicting $\sin\theta_{23}=0$ are excluded.}
\label{table:s13is0}
\end{center}
\end{table}

\medskip

\subsection{Setting $\theta_{13}$ and $\delta$ to the experimental values}
\label{sec:13del_exp}

\begin{figure}[htbp]
\begin{center}
\includegraphics[width=10cm]{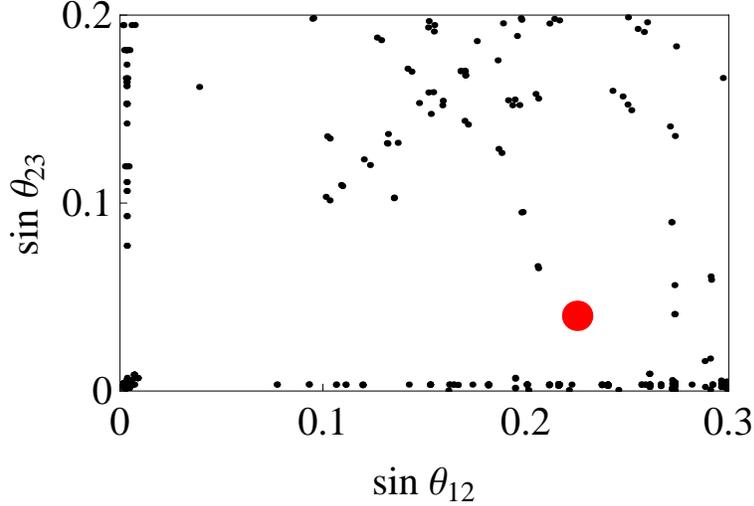}\medskip
\caption{
Legend is the same as Fig. \ref{fig:s13is0}, but for $(\sin\theta_{13},\, \cos\delta)=(0.00341,\, 0.355)$.
}
\label{fig:s13delta_exp}
\end{center}
\end{figure}

In the previous subsections, we set the small CKM angles to zero.
We here consider a more realistic case by setting $\theta_{13}$ and $\delta$ to their experimental values: $(\sin\theta_{13},\, \cos\delta)=(0.00341,\, 0.355)$.
The results up to $(n,\,m,\,p)=(6,\,6,\,12)$ are displayed in Fig. \ref{fig:s13delta_exp}. 
Comparing with Fig. \ref{fig:s13is0}, one can find that some of the solutions appear in both figures.
In addition to such solutions, there also exist solutions which do not appear in Fig. \ref{fig:s13is0}.
Seven solutions close to the experimental values are picked up in Table \ref{table:s13delta_exp}.
It seems that the tendency for large groups to generate realistic
values still holds, while the combination $(p,q_2)=(7,3)$ is included
as one of the simplest ones again.
Nevertheless, as variety has come to the solution, one can find favorable solutions more easily. 
Note that similar results are obtained by fixing two other CKM parameters.

\medskip

\begin{table}[htbp]
\begin{center}
\begin{align}
\begin{array}{c|ccc|cccc|ccc}\hline\hline
(\sin\theta_{12},\,\sin\theta_{23})&n&m & p & k_u&k_c&k_d&k_s&q_1&q_2& (\text{Re}[a],\, \text{Im}[a])\\ \hline
(0.223,\, 0.00341)&4&4&7&1& 3&1&  3&0&3&(-0.802,\, 0)\\
(0.221,\, 0.000522)&4&4&10&1&2& 0& 3&2&3&(-1.00,\, 1.90)\\
(0.273,\, 0.0408) &2&5&12&0&1&2&3&1&5&(-1.00,\, 1.00)\\
(0.273,\, 0.0563) &5&4&10&2&4&1&3&5&7&(-1.00,\, -1.90)\\
(0.271,\, 0.0898) &5&6&10&1& 2& 2&1&5&6&(-1.00,\, -1.18)\\
(0.291,\, 0.0173) &2&6&11&0&1&1&2&0&2&(1.83,\, 0)\\
(0.288,\, 0.0159)&5&6&11&1&4&5&1&0&1&(2.68,\, 0)\\ \hline
\end{array}\nonumber
\end{align}\medskip
\caption{
Legend is the same as Table \ref{table:s13is0}, but for $(\sin\theta_{13},\, \cos\delta)=(0.00341,\, 0.355)$.
}
\label{table:s13delta_exp}
\end{center}
\end{table}

\medskip

\subsection{$m=2$ case}

Lastly, we consider the case of $m=2$ as the situation is slightly different from the other cases. 
In this case, Eq.~\eqref{eq:trW=a} turns out to be
\begin{align}
 |V_{ui}|^2&\,=\,\frac{\text{Re}[a] \cos \frac{\phi_c+\phi_t}{2}+\cos \frac{\phi_{c}+\phi_{t}-2\phi_{u}}{2}+\text{Im}[a] \sin \frac{\phi_c+\phi_t}{2}}{4\sin \frac{\phi_{c}-\phi_u}{2}\sin \frac{\phi_u-\phi_t}{2}}\,, \nonumber \\
  |V_{c i}|^2&\,=\,\frac{\text{Re}[a] \cos \frac{\phi_u+\phi_t}{2}+\cos \frac{\phi_u+\phi_t-2\phi_c}{2}+\text{Im}[a] \sin \frac{\phi_u+\phi_t}{2}}{4\sin \frac{\phi_{u}-\phi_c}{2}\sin \frac{\phi_{c}-\phi_{t}}{2}}\,, \nonumber \\
    |V_{t i}|^2&\,=\,\frac{\text{Re}[a] \cos \frac{\phi_u+\phi_c}{2}+\cos \frac{\phi_u+\phi_c-2\phi_t}{2}+\text{Im}[a] \sin \frac{\phi_u+\phi_c}{2}}{4\sin \frac{\phi_{t}-\phi_{u}}{2}\sin \frac{\phi_{c}-\phi_{t}}{2}} \,, 
    \label{eq:m=2}
\end{align}
as derived in Ref.~\cite{Hernandez:2012ra} for the lepton sector.
As a result, one can directly constrain the $i$th column of the CKM matrix without inputting the CKM parameters by hand.
Note that the unitary condition, $|V_{ui}|^2+ |V_{c i}|^2 +|V_{t i}|^2=1$, has been used, and the subscript $i$ reflects the position of a positive sign in $S_0$, e.g.  $S_0=\text{diag} \{1,\,-1,\,-1\}$ for $i=d$.
In the following, we take $i=s$, for the experimental values of the corresponding column are slightly large. 

\begin{figure}[htbp]
\begin{center}
\includegraphics[width=10cm]{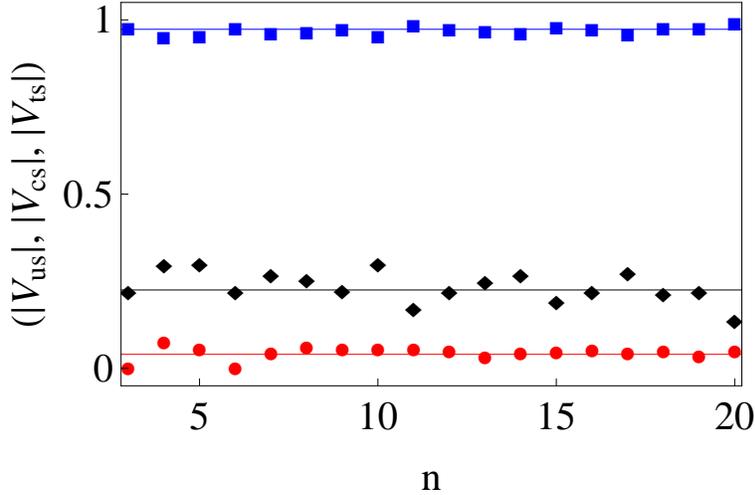}\medskip
\caption{
$|V_{us}|$, $|V_{cs}|$, and $|V_{ts}|$ are plotted as a function of $n$ up to $(n,\,p) = (20,\,20)$ while assuming $m=2$.
The black diamonds, blue squares, and red circles represent $|V_{us}|$, $|V_{cs}|$, and $|V_{ts}|$, respectively.
The horizontal lines describe the experimental values: $(|V_{us}|,\,|V_{cs}|,\,|V_{ts}|)=(0.225,\, 0.973,\, 0.0404)$.
For each $n$, we pick out a solution which derives the nearest mixing to the experimental values. 
}
\label{fig:m=2}
\end{center}
\end{figure}

Results up to $(n,\,p)=(6,\,12)$ are similar to those in Sec. \ref{sec:13del_exp}.
Thus, we here vary $(n,\,p)$ up to $(20,\,20)$ and show that large groups can derive a much more realistic CKM mixing.
For each $n$, in Fig. \ref{fig:m=2}, we pick out a solution which best reproduces the observed CKM mixing. 
Furthermore, we select the seven nearest solutions from Fig. \ref{table:m=2} and list them in Table \ref{table:m=2}. 
As can be seen, one can easily find solutions near the experimental values with large $n$ and $p$. 

\medskip

\begin{table}[htbp]
\begin{center}
\begin{align}
\begin{array}{c|cc|cc|ccc}\hline\hline
(|V_{us}|,\,|V_{cs}|,\,|V_{ts}|)&n & p & k_u&k_c&q_1&q_2& (\text{Re}[a],\, \text{Im}[a])\\ \hline
(0.269,\, 0.962,\, 0.0444)& 7& 19& 0& 2&4&5& (-0.823, 1.80)\\ 
(0.218,\, 0.975,\, 0.0510)&12&19& 3& 8&11&12&(-1.31, -2.18)\\
(0.248,\, 0.968,\, 0.0322)&13&19&5&7&10&11&(-1.08, -1.25)\\
(0.190,\, 0.981,\, 0.0468)&15&13&1&10&7&9&(-1.21, -2.17)\\
(0.275,\, 0.960,\, 0.0444)&17&18&5&11&10&11&(-1.21, -1.85)\\
(0.215,\, 0.975,\, 0.0507)&18&20&6& 11&11&12&(-1.17, -1.71)\\
(0.219,\, 0.975,\, 0.0345)&19&15&6& 1&1&2&(1.89, 0.199)\\
\hline
\end{array}\nonumber
\end{align}\medskip
\caption{From Fig. \ref{fig:m=2}, we pick out seven solutions close to the experimental values.}
\label{table:m=2}
\end{center}
\end{table}

\medskip

\section{EMBEDDING INTO $\Delta(6N^2)$}\label{sec:delta}

We have considered the von Dyck group. 
Constraining to some special cases, it is possible to embed 
the obtained residual symmetries into a finite group, for example, $\Delta (6N^2)$. 
We stress that this group is also useful to 
provide the mixing angles of leptons 
under the experimentally allowed region.

\begin{figure}[htbp]
\begin{center}
\includegraphics[width=10cm]{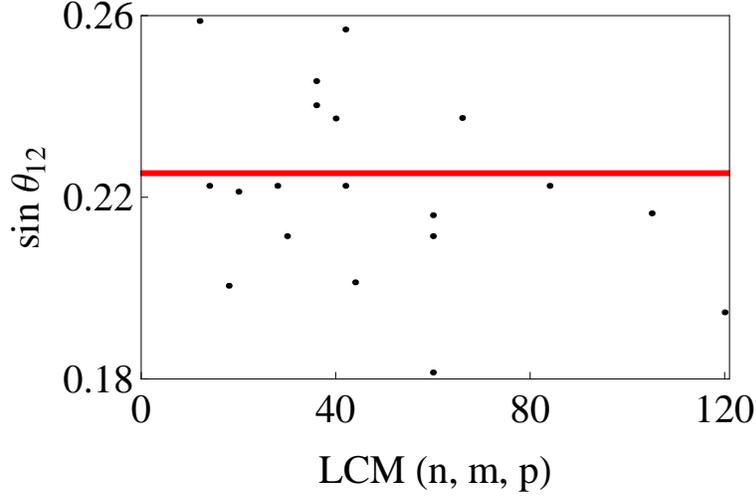}\medskip
\caption{Distribution of $\sin\theta_{12}$ as a function of the least common multiples of $(n,\,m,\,p)$ up to $(6,\,6,\,12)$ in the case of $\theta_{13}=\theta_{23}=0$.}
\label{fig:cab2}
\end{center}
\end{figure}

In this section, we focus on the case with $\theta_{13}=\theta_{23}=0$ 
as studied in Sec. \ref{sec:cabibbo}. 
For the purpose of embedding residual symmetries into a smaller $\Delta (6N^2)$ series, 
we redraw the plots in Fig. \ref{fig:cab} 
as a function of the least common multiples (LCMs) of $(n,\, m,\,p)$ to Fig. \ref{fig:cab2}.

\medskip

For example, 
in Figure \ref{fig:cab2}, the solution with $\text{LCM}(n,\,m,\,p)=12$ corresponds to
\begin{equation}
(\sin\theta_{12},\,n,\,m,\,p)\,=\,(0.259,\,2,\,2,\,12)\,,
\end{equation}
which is also found in Table I of Ref.~\cite{HolLim}.
This solution can be derived by $\Delta (6\times 12^2)$ 
and it realizes the value by 
$\sin\theta_{12}=\frac{1}{2} |e^{2\pi i/12}+e^{\pi i}|$. 
The detail of this derivation is explained later. 
Another example is $\text{LCM}(n,\,m,\,p)=14$ 
with $\sin\theta_{12}=0.223$, 
and we can derive it by $\Delta (6\times 7^2)$ 
with the value of 
$\sin\theta_{12}=\frac{1}{2} |e^{2\pi i/7}+e^{8\pi i/7}|$.

At first, let us simply review the group theory of $\Delta(6N^2)$~\cite{Escobar:2008vc,Ishimori:2010au}. 
Elements of this group are given by the products of $a$, $a'$, $b$, and $c$ satisfying $a^N=a'^N=b^3=c^2=e$.
In addition, they obey $ba^xa'^y b^2=a^{-x+y}a'^{-x}$, $b^2a^xa'^y b=a^{-y}a'^{x-y}$, and $cb^zc=b^{2z}$. 
All the group elements are expressed by $g=b^zc^w a^x a'^y$. 

To derive the Cabibbo angle, we use the $Z_2$ elements of $\Delta(6N^2)$, namely, elements satisfying $g^2=e$. 
We can write down $Z_2$ elements for fixed $z$ and $w$. 
For $(z,\,w)=(0,\,0)$, $(1,\,0)$, and $(2,\,0)$, $g^2$ is given as $a^{2x}a'^{2y}$, $b^2a^{x-y}a'^{x}$, and $ba^{y}a'^{y-x}$, respectively. 
Then $Z_2$ elements exist in $(z,\,w)=(0,\,0)$, which are $a^{N/2}$, $a'^{N/2}$, and $a^{N/2}a'^{N/2}$ when $N$ is even. 
Similarly, for $(z,w)=(0,1)$, $(1,1)$, and $(2,1)$, $g^2$ is given as $g^2=a^{x-y}a'^{-x+y}$, $a^{y}a'^{2y}$, and $a^{2x}a'^{x}$, respectively. 
Then $Z_2$ elements are written by $ca^xa'^x$, $bca^x$, and $b^2ca'^y$.
 
Here we identify $Z_2$ symmetries in down and up quark mass terms with $bca^{x}$ and $bca^{y}$, respectively. 
In the diagonalizing basis of the up quark mass terms, matrix representations are expressed as
\begin{align}
S_D\,=\,
\begin{pmatrix}
\cos\left[\frac{2\pi}{N}(x-y)\right]&&i \sin\left[\frac{2\pi}{N}(x-y)\right]&&0\\[1ex]
-i \sin\left[\frac{2\pi}{N}(x-y)\right]&&-\cos\left[\frac{2\pi}{N}(x-y)\right]&&0\\[1ex]
0&&0&&1
\end{pmatrix}\,,
\end{align}
and $T=\text{diag}\{1,\,-1,\,-1\}$.
Setting $S_0=\text{diag}\{-1,\,1,\,-1\}$, the CKM matrix is obtained as
\begin{align}
V_\text{CKM}\,=\,
\begin{pmatrix}
(1-\rho^{x-y})/2&&(1+\rho^{x-y})/2&&0\\
(1+\rho^{x-y})/2&&(1-\rho^{x-y})/2&&0\\
0&&0&&1
\end{pmatrix}\,,
\end{align}
where $\rho=e^{2\pi i/N}$.
Since the overall phase is irrelevant, 
we obtain $\sin\theta_{12}=\cos\left[\pi (x-y)/N\right]$. 
This formula is the same as the one derived in Sec. \ref{sec:cabibbo} 
i.e.,  $\sin\theta_{12}=\cos(\pi q_2/p)$. 
When $N=7$ and $x-y=3$, it yields $|V_{us}|=0.223$, which
is very close to the experimental value. 
As commented in Sec. \ref{sec:cabibbo}, 
there is no integer $N$ for $N < 89$, which fits better than 
$N=7$ except $(N,x-y)=(7r,3r)$.
Thus, the $\Delta(6N^2)$ with $N=7r$ seems favorable.
Figure 6 shows $\sin \theta_{12}$ for several values of $N$ and $x-y$.

\begin{figure}[htbp]
\begin{center}
\includegraphics[width=10cm]{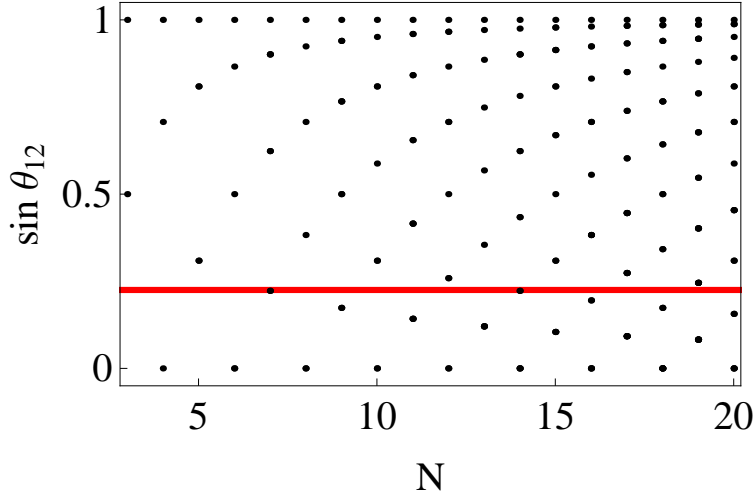}\medskip
\caption{$\sin\theta_{12}$ versus $N$ are plotted. 
$N$ is the number of $\Delta(6N^2)$ 
that is taken from $3$ to $20$. The continuous line 
describes the experimental value, $\sin\theta_{12}=0.225$.}
\end{center}
\end{figure}

Next, we consider mixing angles of the lepton sector 
by using other group elements. 
We find that $Z_3$ elements are useful to explain large 
mixing angles of leptons. 
Explicitly, $Z_3$ elements 
of the group are $a^{iN/3}a'^{jN/3}$, 
$ba^xa'^y$, and $b^2a^xa'^y$ where 
$i,j=0,1,2$, and $N/3$ is assumed to be the integer. 
This time, we take the charged lepton $Z_3$ symmetry as $T'\equiv ba^{-x}a'^{-x-y}$. 
This representation can be written and diagonalized by
\begin{align}T'=
\begin{pmatrix}
0&\rho^{-y}&0\\
0&0&\rho^{x+y}\\
\rho^{-x}&0&0
\end{pmatrix}
=U_{T'}
\begin{pmatrix}
1&0&0\\
0&\omega&0\\
0&0&\omega^2
\end{pmatrix}
U_{T'}^\dagger \,,
\quad
U_{T'}
=\frac{1}{\sqrt3}
\begin{pmatrix}
\rho^{x}&\omega\rho^x&\omega^2\rho^x\\
\rho^{x+y}&\omega^2\rho^{x+y}&\omega\rho^{x+y}\\
1&1&1\\
\end{pmatrix}\,,
\end{align}
where $\omega=e^{2\pi i/3}$.
For the neutrino $Z_2$ symmetry, we choose $S'\equiv b^2ca'^{x'}$ so that
\begin{align}S'=
\begin{pmatrix}
-1&0&0\\
0&0&-\rho^{-x'}\\
0&-\rho^{x'}&0
\end{pmatrix}
=U_{S'}
\begin{pmatrix}
1&0&0\\
0&-1&0\\
0&0&-1
\end{pmatrix}
U_{S'}^\dagger \,,
\quad
U_{S'}
=\begin{pmatrix}
0&1&0\\
-\rho^{-x'}/\sqrt2&0& 1/\sqrt2\\
1/\sqrt2&0&\rho^{x'}/\sqrt2\\
\end{pmatrix}\,.
\end{align}
Then the mixing matrix of the lepton sector becomes
\begin{align}
U_\text{PMNS}
\,=\,U_{T'}^\dagger U_{S'} \,=\,\begin{pmatrix}
\frac{1-\rho^{-x-y-x'}}{\sqrt6}
&\frac{\rho^{-x}}{\sqrt3}
&\frac{\rho^{-x-y}+\rho^{x'}}{\sqrt6}
\\
\frac{1-\omega\rho^{-x-y-x'}}{\sqrt6}
&\frac{\omega^2\rho^{-x}}{\sqrt3}
&\frac{\omega\rho^{-x-y}+\rho^{x'}}{\sqrt6}
\\
\frac{1-\omega^2\rho^{-x-y-x'}}{\sqrt6}
&\frac{\omega\rho^{-x}}{\sqrt3}
&\frac{\omega^2\rho^{-x-y}+\rho^{x'}}{\sqrt6}
\end{pmatrix}.
\end{align}

The same mixing matrix has been analyzed by~\cite{Ishimori:2012gv,King:2013vna} 
and it predicts the sum rules $\theta_{23}\approx45^\circ\mp \theta_{13}/\sqrt2$. 
Also, it realizes the trimaximal mixing $\sin^2\theta_{12}\approx 1/3$ for small $\theta_{13}$. 
In particular, since $\sin\theta_{13}$ can be written as 
$\sqrt2 \cos(\pi(x+y+x')/N)/\sqrt3$,  small $\theta_{13}$ 
is realized by $(x+y+x')/N\approx 1/2$. 
When $N=7$ and $x+y+x'=3$, we have 
$\sin^22\theta_{13}=0.128$ which is close to the experimental value. 
As a result, the $\Delta(6N^2)$ for $N=7r$ is interesting 
to realize the mixing angles for both the quark and lepton sectors.

\medskip

\section{SUMMARY}\label{sec:summary}

We have applied the model-independent formalism, which was recently developed by Hernandez and Smirnov, to the quark sector and sought possible residual symmetries with a focus on the von Dyck groups.
In the case of $\theta_{13}=\theta_{23}=0$, the Cabibbo angle
$\theta_{12}$ can be close to its experimental values for small $n$,
$m$, and $p$.
In particular, the combination between the $Z_2$ and $Z_7$ symmetries seems favorable
to realize the realistic value of $\theta_{12}$.
Furthermore, these residual symmetries can originate from finite
groups such as $D_{N}$ and $\Delta(6N^2)$ with $N=7r$.    

Indeed, we have also discussed possibilities of embedding the obtained
residual symmetries into the $\Delta(6N^2)$ series.
It is found that $\Delta(6N^2)$ for $N=7r$ would be favorable 
to realize the Cabibbo angle 
and also interesting from the viewpoint of the mixing angles 
for the lepton sector. 

In contrast, relatively large $n$, $m$, and $p$ are needed in order to
reproduce all the quark mixing parameters at the same time.
The von Dyck groups with such large integers correspond to 
not finite groups, but infinite ones.
However, they may be embedded into finite subgroups of infinite
von Dyck groups (see e.g. Ref.~\cite{deAdelhartToorop:2011re}).
The work is in progress, and we would like to postpone this issue to
our next publication.
At any rate, we have shown which combinations of 
residual symmetries lead to favorable results.
That can become a starting point to investigate the full flavor symmetry 
hiding behind the quark and lepton mass matrices.

\bigskip

\section*{Acknowledgement}

This work was supported in part by the Grant-in-Aid for Scientific Research 
No. 23.696 (H.Ishimori), No.~25400252 (T.K.) and No. 25.1146 (A.O.) from the Ministry of Education, Culture, Sports, Science 
and Technology of Japan. H.Ishida was financially supported partly by Inoue Foundation for Science.


\begin{thebibliography}{99}

\bibitem{pdg}
J. Beringer et al. (Particle Data Group), 
Phys.\ Rev.\ D {\bf 86}, 010001 (2012).


\bibitem{Harrison:2002er}
  P.~F.~Harrison, D.~H.~Perkins and W.~G.~Scott,
  %``Tri-bimaximal mixing and the neutrino oscillation data,''
  Phys.\ Lett.\  B {\bf 530}, 167 (2002)
  [arXiv:hep-ph/0202074];
  %%CITATION = PHLTA,B530,167;%%
%
  Z.~Z.~Xing, 
  Phys.\ Lett.\ B {\bf 533}, 85 (2002) 
  [arXiv:hep-ph/0204049]; 
%
%
%
%\bibitem{Harrison:2002kp}
  P.~F.~Harrison and W.~G.~Scott,
  %``Symmetries and generalisations of tri-bimaximal neutrino mixing,''
  Phys.\ Lett.\  B {\bf 535}, 163 (2002)
  [arXiv:hep-ph/0203209];
  %%CITATION = PHLTA,B535,163;%%
%
%\cite{Harrison:2003aw}
%\bibitem{Harrison:2003aw}
%  P.~F.~Harrison and W.~G.~Scott,
  %``Permutation symmetry, tri-bimaximal neutrino mixing and the S3 group
  %characters,''
  Phys.\ Lett.\  B {\bf 557}, 76 (2003)
  [arXiv:hep-ph/0302025].
  %%CITATION = PHLTA,B557,76;%%


%\cite{Altarelli:2010gt}
\bibitem{Altarelli:2010gt} 
  G.~Altarelli and F.~Feruglio,
  %``Discrete Flavor Symmetries and Models of Neutrino Mixing,''
  Rev.\ Mod.\ Phys.\  {\bf 82}, 2701 (2010)
  [arXiv:1002.0211 [hep-ph]].
  %%CITATION = ARXIV:1002.0211;%%
 

%\cite{Ishimori:2010au}
\bibitem{Ishimori:2010au} 
  H.~Ishimori, T.~Kobayashi, H.~Ohki, Y.~Shimizu, H.~Okada and M.~Tanimoto,
  %``Non-Abelian Discrete Symmetries in Particle Physics,''
  Prog.\ Theor.\ Phys.\ Suppl.\  {\bf 183}, 1 (2010)
  [arXiv:1003.3552 [hep-th]];
  %%CITATION = ARXIV:1003.3552;%%
%\cite{Ishimori:2012zz}
%\bibitem{Ishimori:2012zz} 
%  H.~Ishimori, T.~Kobayashi, H.~Ohki, H.~Okada, Y.~Shimizu and M.~Tanimoto,
  %``An introduction to non-Abelian discrete symmetries for particle physicists,''
  Lect.\ Notes Phys.\  {\bf 858}, 1 (2012);
  %%CITATION = LNPHA,858,pp.1;%%
%\cite{Ishimori:2013woa}
%\bibitem{Ishimori:2013woa} 
%  H.~Ishimori, T.~Kobayashi, Y.~Shimizu, H.~Ohki, H.~Okada and M.~Tanimoto,
  %``Non-Abelian discrete symmetry for flavors,''
  Fortsch.\ Phys.\  {\bf 61}, 441 (2013).
  %%CITATION = FPYKA,61,441;%%


%\cite{King:2013eh}
\bibitem{King:2013eh} 
  S.~F.~King and C.~Luhn,
  %``Neutrino Mass and Mixing with Discrete Symmetry,''
  Rep.\ Prog.\ Phys.\  {\bf 76}, 056201 (2013)
  [arXiv:1301.1340 [hep-ph]].
  %%CITATION = ARXIV:1301.1340;%%


\bibitem{resdS4}
C. S. Lam, 
Phys. Lett. B{\bf 656}, 193 (2007) 
[arXiv:0708.3665 [hep-ph]];
Phys. Rev. Lett. {\bf 101}, 121602 (2008)
[arXiv:0804.2622 [hep-ph]];
Phys. Rev. D{\bf 78}, 073015 (2008)
[arXiv:0809.1185 [hep-ph]];
W. Grimus, L. Lavoura and P. O. Ludl,
J. Phys. G{\bf 36}, 115007 (2009)
[arXiv:0906.2689 [hep-ph]].


\bibitem{rct}
DAYA-BAY Collaboration, 
Phys.\ Rev.\ Lett.\ {\bf 108}, 171803 (2012)
[arXiv:1203.1669 [hep-ex]];
Chin.\ Phys.\ C {\bf 37}, 011001 (2013)
[arXiv:1210.6327 [hep-ex]];
DOUBLE-CHOOZ Collaboration, 
Phys.\ Rev.\ Lett.\ {\bf 108}, 131801 (2012)
[arXiv:1112.6353 [hep-ex]];
RENO Collaboration, 
Phys.\ Rev.\ Lett.\ {\bf 108}, 191802 (2012)
[arXiv:1204.0626 [hep-ex]].

\bibitem{acl}
T2K Collaboration, 
Phys.\ Rev.\ Lett.\ {\bf 107}, 041801 (2011)
[arXiv:1106.2822 [hep-ex]];
MINOS Collaboration, 
Phys.\ Rev.\ Lett.\ {\bf 107}, 181802 (2011)
[arXiv:1108.0015 [hep-ex]].

%\cite{Ge:2011qn}
\bibitem{Ge:2011qn} 
  S.~-F.~Ge, D.~A.~Dicus and W.~W.~Repko,
  %``Residual Symmetries for Neutrino Mixing with a Large $\theta_{13}$ and Nearly Maximal $\delta_D$,''
  Phys.\ Rev.\ Lett.\  {\bf 108}, 041801 (2012)
  [arXiv:1108.0964 [hep-ph]];
  %%CITATION = ARXIV:1108.0964;%%
%\cite{Ge:2011ih}
%\bibitem{Ge:2011ih} 
 %S.~-F.~Ge, D.~A.~Dicus and W.~W.~Repko,
  %``Z_2 Symmetry Prediction for the Leptonic Dirac CP Phase,''
  Phys.\ Lett.\ B {\bf 702}, 220 (2011)
  [arXiv:1104.0602 [hep-ph]].
  %%CITATION = ARXIV:1104.0602;%%

%\cite{Hernandez:2012ra}
\bibitem{Hernandez:2012ra} 
  D.~Hernandez and A.~Y.~Smirnov,
  %``Lepton mixing and discrete symmetries,''
  Phys.\ Rev.\ D {\bf 86}, 053014 (2012)
  [arXiv:1204.0445 [hep-ph]];
  %%CITATION = ARXIV:1204.0445;%%
%
%\cite{Hernandez:2012sk}
%\bibitem{Hernandez:2012sk} 
%  D.~Hernandez and A.~Y.~.Smirnov,
  %``Discrete symmetries and model-independent patterns of lepton mixing,''
  Phys.\ Rev.\ D {\bf 87}, 053005 (2013)
  [arXiv:1212.2149 [hep-ph]];
  %%CITATION = ARXIV:1212.2149;%%
%
  arXiv:1304.7738 [hep-ph].
  
   %\cite{Hu:2012ei}
\bibitem{Hu:2012ei} 
  B.~Hu,
  %``Neutrino Mixing and Discrete Symmetries,''
  Phys.\ Rev.\ D {\bf 87}, 033002 (2013)
  [arXiv:1212.2819 [hep-ph]];
  %%CITATION = ARXIV:1212.2819;%%
%\cite{Lam:2013xs}
%\bibitem{Lam:2013xs} 
  C.~S.~Lam,
  %``Leptonic Mixing and Group Structure Constants,''
  Phys.\ Rev.\ D {\bf 87}, 053018 (2013)
  [arXiv:1301.3121 [hep-ph]];
  %%CITATION = ARXIV:1301.3121;%%
%\cite{Ballett:2013wya}
%\bibitem{Ballett:2013wya} 
  P.~Ballett, S.~F.~King, C.~Luhn, S.~Pascoli and M.~A.~Schmidt,
  %``Testing atmospheric mixing sum rules at precision neutrino facilities,''
  arXiv:1308.4314 [hep-ph].
  %%CITATION = ARXIV:1308.4314;%%

%\cite{Araki:2008ek}
\bibitem{Araki:2008ek} 
  T.~Araki,
  %``Anomaly of Discrete Symmetries and Gauge Coupling Unification,''
  Prog.\ Theor.\ Phys.\  {\bf 117}, 1119 (2007)
  [arXiv:hep-ph/0612306];
  %%CITATION = HEP-PH/0612306;%%
  T.~Araki, T.~Kobayashi, J.~Kubo, S.~Ramos-Sanchez, M.~Ratz and P.~K.~S.~Vaudrevange,
  %``(Non-)Abelian discrete anomalies,''
  Nucl.\ Phys.\ B {\bf 805}, 124 (2008)
  [arXiv:0805.0207 [hep-th]];
  %%CITATION = ARXIV:0805.0207;%%
%\cite{Luhn:2008sa}
%\bibitem{Luhn:2008sa} 
  C.~Luhn and P.~Ramond,
  %``Anomaly Conditions for Non-Abelian Finite Family Symmetries,''
  JHEP {\bf 0807}, 085 (2008)
  [arXiv:0805.1736 [hep-ph]].
  %%CITATION = ARXIV:0805.1736;%%

\bibitem{D7}
  A.~Blum, C.~Hagedorn and M.~Lindner,
  Phys.\ Rev.\ D {\bf 77}, 076004 (2008)
  [arXiv:0709.3450 [hep-ph]];
%
  A.~Blum, C.~Hagedorn and A.~Hohenegger,
  JHEP {\bf 0803}, 070 (2008)
  [arXiv:0710.5061 [hep-ph]];
%\cite{Hagedorn:2012pg}
%\bibitem{Hagedorn:2012pg} 
  C.~Hagedorn and D.~Meloni,
  %``D14 - A Common Origin of the Cabibbo Angle and the Lepton Mixing Angle theta^l_13,''
  Nucl.\ Phys.\ B {\bf 862}, 691 (2012)
  [arXiv:1204.0715 [hep-ph]].
  %%CITATION = ARXIV:1204.0715;%%


\bibitem{deAdelhartToorop:2011re} 
  R.~de Adelhart Toorop, F.~Feruglio and C.~Hagedorn,
  %``Finite Modular Groups and Lepton Mixing,''
  Nucl.\ Phys.\ B {\bf 858}, 437 (2012)
  [arXiv:1112.1340 [hep-ph]].
  %%CITATION = ARXIV:1112.1340;%%



%\cite{Escobar:2008vc}
\bibitem{Escobar:2008vc} 
  J.~A.~Escobar and C.~Luhn,
  %``The Flavor Group Delta(6n**2),''
  J.\ Math.\ Phys.\  {\bf 50}, 013524 (2009)
  [arXiv:0809.0639 [hep-th]].
  %%CITATION = ARXIV:0809.0639;%%


\bibitem{HolLim}
  M.~Holthausen and K.~S.~Lim,
  Phys.\ Rev.\ D {\bf 88}, 033018 (2013)
  [arXiv:1306.4356 [hep-ph]].
  
\bibitem{Ishimori:2012gv} 
  H.~Ishimori and T.~Kobayashi,
  %``Lepton flavor models with discrete prediction of theta_{13},''
  Phys.\ Rev.\ D {\bf 85}, 125004 (2012)
  [arXiv:1201.3429 [hep-ph]].
  %%CITATION = ARXIV:1201.3429;%%
 


%\cite{King:2013vna}
\bibitem{King:2013vna} 
  S.~F.~King, T.~Neder and A.~J.~Stuart,
  %``Lepton Mixing Predictions from Delta(6n^2) Family Symmetry,''
  Phys.\ Lett.\ B {\bf 726}, 312 (2013)
  [arXiv:1305.3200 [hep-ph]].
  %%CITATION = ARXIV:1305.3200;%%



\end{thebibliography}
\end{document}